\documentclass[11pt,twoside]{article}
\usepackage{asp2014}

\aspSuppressVolSlug
\resetcounters

\begin{document}

\title{The Gamma-Flash data acquisition system for observation of terrestrial gamma-ray flashes}

\author{A.~Bulgarelli$^1$, 
A.~Addis$^1$, 
A.~Aboudan$^{1, 2}$, 
I.~Abu$^1$, 
C.~Andreani$^3$, 
A.~Argan$^4$, 
R.~Campana$^1$, 
P.~Calabretto$^1$, 
C.~Pittori$^{5,6}$, 
F.~D'Amico$^7$, 
I.~Donnarumma$^7$, 
A.~De Rosa$^1$, 
F.~Fuschino$^1$, 
G.~Gorini$^8$, 
G.~Levi$^{1,9}$, 
N.~Parmiggiani$^1$, 
P.~Picozza$^3$, 
G.~Polenta$^7$, 
E.~Preziosi$^3$, 
R.~Senesi$^3$, 
A.~Ursi$^4$, 
V.~Vagelli$^7$, 
E.~Virgilli$^1$}
\affil{$^1$INAF-OAS, Bologna, Via Gobetti 93/3, Bologna, Italy}
\affil{$^2$University of Padova, CISAS, Padova, Italy}
\affil{$^3$Physics Department and NAST Centre University of Rome "Tor Vergata" Via della Ricerca Scientifica 1, 00133 Roma, Italy}

\affil{$^4$INAF-IAPS Roma, Via del Fosso del Cavaliere 100, 00133 Roma, Italy.}
\affil{$^5$INAF-OAR, via Frascati 33, Monte Porzio Catone (RM), Italy}
\affil{$^6$SSDC-ASI, via del Politecnico snc, I-00133 Roma, Italy}
\affil{$^7$ASI, Via del Politecnico snc, 00133 Roma, Italy.}

\affil{$^8$Dip. di Fisica “G. Occhialini,” Università  Milano-Bicocca,  Milan, Italy}

\affil{$^9$DIFA, Dipartimento di Fisica, Università di Bologna, Italy}

\paperauthor{A. Addis}{antonio.addis@inaf.it}{0000-0002-0886-8045}{INAF}{OAS}{Bologna}{Emilia-Romagna}{40129}{Italy}
\paperauthor{A. Bulgarelli}{andrea.bulgarelli@inaf.it}{0000-0001-6347-0649}{INAF}{OAS}{Bologna}{Emilia-Romagna}{4019}{Italy}
\paperauthor{N. Parmiggiani}{nicolo.parmiggiani@inaf.it}{0000-0002-4535-5329}{INAF}{OAS}{Bologna}{Emilia-Romagna}{4019}{Italy}




  
\begin{abstract}

Gamma-Flash is an Italian project funded by the Italian Space Agency (ASI) and led by the National Institute for Astrophysics (INAF), devoted to the observation and study of high-energy phenomena, such as terrestrial gamma-ray flashes and gamma-ray glows produced in the Earth's atmosphere during thunderstorms. The project's detectors and the data acquisition and control system (DACS) are placed at the "O. Vittori" observatory on the top of Mt. Cimone (Italy). A payload will be placed on an aircraft to observe thunderstorms in the air. This work presents the architecture of the data acquisition and control system and the data flow.
  
\end{abstract}

\section{Introduction}

Gamma-Flash is an Italian project funded by the Italian Space Agency (ASI) and led by the National Institute for Astrophysics (INAF) \citep{2022RemS...14.3103U}, devoted to the observation and study of both short-duration transients, such as terrestrial gamma-ray flashes (TGFs), as well as longer-lasting gamma-ray emissions, such as gamma-ray glows, and associated high-energy particles (e.g. neutron) emissions. The program's primary targets are studying thunderstorm-related high-energy emissions, which can substantially impact many fields, such as local/global climate change, environmental studies, and atmospheric plasma physics. TGFs were first discovered in 1994 by BATSE \cite{doi:10.1126/science.264.5163.1313} and then observed by
NASA RHESSI, Fermi, and ASI AGILE satellites in the high-energy band. The Gamma-Flash experiment is a heritage of the AGILE satellite \citep{PhysRevLett.106.018501r}.
The TGFs detected by AGILE/MCAL was presented in the 3rd AGILE TGF Catalog \cite{https://doi.org/10.1029/2019JD031985, https://doi.org/10.1029/2019JD031986}; the updated online version is publicly accessible through a dedicated SSDC
website\footnote{https://www.ssdc.asi.it/mcal3tgfcat/}.

The Gamma-Flash 
project has developed an innovative gamma-ray and neutron detector system and correlated instruments  to be placed on-ground at the Climatic Observatory "O. Vittori" (see Figure 1A and 1B) on the top of Mt. Cimone (2165 m a.s.l., Northern-Central Italy) \citep{rs14143501}. A payload is developed to be placed on aircraft to observe thunderstorms in the air (see Figure 1C).

\articlefigure{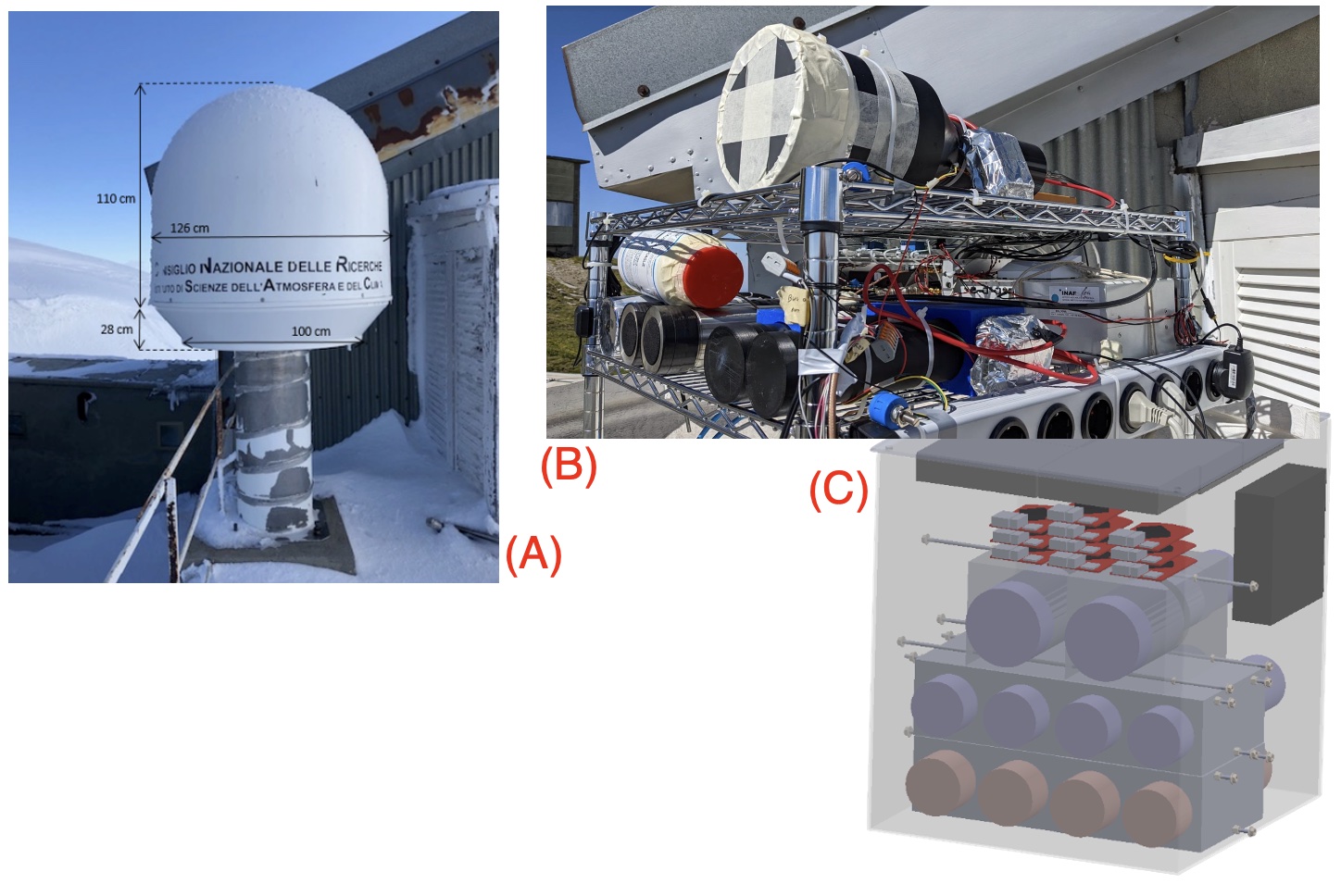}{fig1}{The Gamma-Flash experiment. (A) The dome where the  detectors are placed. (B) Detail of the terrestrial setup, with five $\gamma$-ray  detectors and three neutron detectors. (C)  A model of the Gamma-Flash airborne payload on aircraft flying near the thunderstorm to collect information from a nearby position. \label{fig1}}

\section{Data Acquisition and Control}

The Gamma-Flash Data Acquisition and Control System (DACS) is the Gamma-Flash project's control, archiving, and data processing pipeline (see Figure 2).

\articlefigure{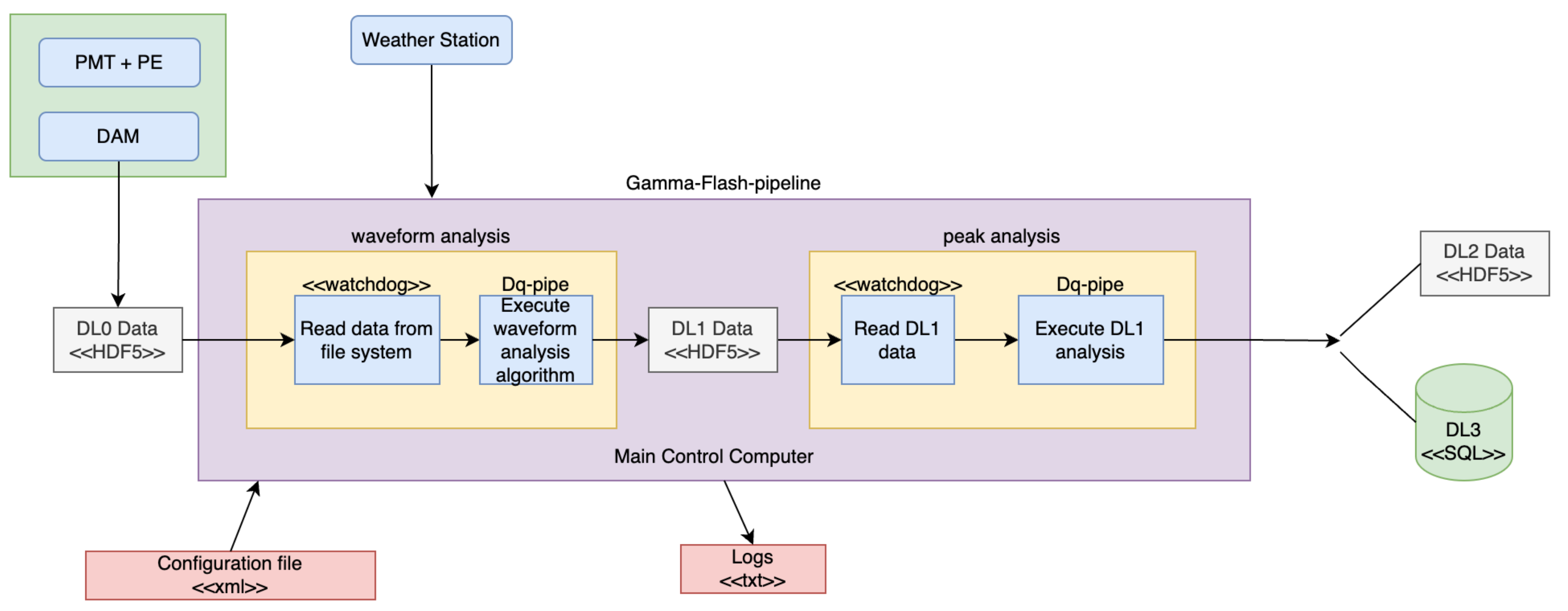}{fig1}{The Data acquisition pipeline of the Gamma-Flash Data Acquisition and Control System (DACS).}

The DACS architecture is modular, and it is composed by:
\begin{itemize}
\item A Photo-multiplier tube detector (PMT) for gamma-ray or neutron events with their proximity electronics (PE);
\item A Data Acquisition Module (DAM) containing a complete data acquisition pipeline made up of an analogue-to-digital converter (ADC), an event triggering logic, and a micro-controller used to handle the data acquisition;
\item The Main Control Computer (MCC). It controls the overall data acquisition process providing data storage, online data processing with the Gamma-Flash pipeline, quick-look capabilities, and a unique interface for remote control of the experiment;
\item A weather station that monitors the environmental temperature and triggers alerts when a specific threshold is reached to provide alarms in real-time or near real-time to the control system.
\end{itemize}

\subsection{Data Acquisition Module}

The Data Acquisition Module (DAM) software comprises two parts: the FPGA firmware (FW) and the application software, both running on a RedPitaya board\footnote{https://redpitaya.com/}.
The DAM controls the waveform acquisition process and implements a low-level TCP socket to send data to the MCC.

\subsection{Main Control Computer}

The Main Control Computer (MCC) is an x86 multi-core machine to receive, processes, and archives the data in different Data Layers (DL):
\begin{itemize}
\item DL0 is an HDF5 file containing raw waveforms from DAMs.
\item DL1 is an HDF5 or Pickle file containing a reduced version of the DL0. DL1 data are obtained from DL0 by applying a filter that only selects peaks due to the inorganic scintillator and which therefore have a duration of $400 ns$ or more.
\item DL2 is an HDF5 file containing an event list computed according to a specific algorithm selected by the scientists.
\item DL3 is a logical data level containing cumulative spectrum histograms and light curves.
\end{itemize}
DL0 is produced by a Python module when a selected number of waveforms are received from DAM. 
The DL1, DL2 and DL3 are generated by the Gamma-Flash pipeline.
The MCC also collects the data from a lightning antenna (480 km detection capability) LD-250 to correlate flashes of lightning with TGFs. 

The Gamma-Flash pipeline produces almost 40 GB of DL0 raw waveforms per day for each detector, reduced to 2 GB at the DL1 data level.

\section{ASI/SSDC Data Access}

The data is transferred from the Monte Cimone site to the INAF/OAS Bologna as temporary storage and from Bologna to ASI Space Science Data Center (SSDC) in Rome, where the final data archiving and data access is provided. 
SSDC will also provide public access to the database of TGFs detected by the Gamma-Flash experiment to the scientific team.

\section{Conclusions}

The Gamma-Flash Data Acquisition and Control System is designed to provide (i)  the interfaces to the remote control to configure and control the experiment; (ii)  on-site data acquisition, monitoring and control capabilities; (iii)  alarms, and rate counting in real-time or near real-time to the remote control; (iv)  the results of the data acquisition through a quick-look dashboard. Final data archiving and access are provided by ASI/SSDC.

\section*{Acknowledgments}
Gamma-Flash is an experiment supported by the ASI-INAF agreement N. 2020-5-HH.0. Gamma-Flash also involves ISAC, University of Rome Tor Vergata, University of Padua, CIFS and INFN.


\bibliography{P42}

\begin{thebibliography}{}
\expandafter\ifx\csname natexlab\endcsname\relax\def\natexlab#1{#1}\fi
\expandafter\ifx\csname url\endcsname\relax
  \def\url#1{\texttt{#1}}\fi
\expandafter\ifx\csname urlprefix\endcsname\relax\def\urlprefix{URL }\fi
\providecommand{\eprint}[2][]{\url{#2}}

\bibitem[{Fishman et~al.(1994)Fishman, Bhat, Mallozzi, Horack, Koshut,
  Kouveliotou, Pendleton, Meegan, Wilson, Paciesas, Goodman, \&
  Christian}]{doi:10.1126/science.264.5163.1313}
Fishman, G.~J., Bhat, P.~N., Mallozzi, R., Horack, J.~M., Koshut, T.,
  Kouveliotou, C., Pendleton, G.~N., Meegan, C.~A., Wilson, R.~B., Paciesas,
  W.~S., Goodman, S.~J., \& Christian, H.~J. 1994, Science, 264, 1313

\bibitem[{Lindanger et~al.(2020)Lindanger, Marisaldi, Maiorana, Sarria,
  Albrechtsen, Østgaard, Galli, Ursi, Labanti, Tavani, Pittori, \&
  Verrecchia}]{https://doi.org/10.1029/2019JD031985}
Lindanger, A., Marisaldi, M., Maiorana, C., Sarria, D., Albrechtsen, K.,
  Østgaard, N., Galli, M., Ursi, A., Labanti, C., Tavani, M., Pittori, C., \&
  Verrecchia, F. 2020, Journal of Geophysical Research: Atmospheres, 125,
  e2019JD031985

\bibitem[{Maiorana et~al.(2020)Maiorana, Marisaldi, Lindanger, Østgaard, Ursi,
  Sarria, Galli, Labanti, Tavani, Pittori, \&
  Verrecchia}]{https://doi.org/10.1029/2019JD031986}
Maiorana, C., Marisaldi, M., Lindanger, A., Østgaard, N., Ursi, A., Sarria,
  D., Galli, M., Labanti, C., Tavani, M., Pittori, C., \& Verrecchia, F. 2020,
  Journal of Geophysical Research: Atmospheres, 125, e2019JD031986

\bibitem[{Tavani et~al.(2011)Tavani, Marisaldi, \& the
  AGILE~Team.}]{PhysRevLett.106.018501r}
Tavani, M., Marisaldi, \& the AGILE~Team. (AGILE Team) 2011, Phys. Rev. Lett.,
  106, 018501

\bibitem[{Tiberia et~al.(2022)Tiberia, Arnone, Ursi, Fuschino, Virgilli,
  Preziosi, Tavani, \& Dietrich}]{rs14143501}
Tiberia, A., Arnone, E., Ursi, A., Fuschino, F., Virgilli, E., Preziosi, E.,
  Tavani, M., \& Dietrich, S. 2022, Remote Sensing, 14

\bibitem[{{Ursi} et~al.(2022){Ursi}, {Rodriguez Fernandez}, {Tiberia},
  {Virgilli}, {Arnone}, {Preziosi}, {Campana}, \&
  {Tavani}}]{2022RemS...14.3103U}
{Ursi}, A., {Rodriguez Fernandez}, G., {Tiberia}, A., {Virgilli}, E., {Arnone},
  E., {Preziosi}, E., {Campana}, R., \& {Tavani}, M. 2022, Remote Sensing, 14,
  3103

\end{thebibliography}


\end{document}